\documentclass[epj,epsf,psfig]{svjour}
\pdfoutput=1
\usepackage{amsmath}

\newcommand{\beq}{\begin{equation}}
\newcommand{\eeq}{\end{equation}}

\newcommand{\pythia}    {{\sc{pythia}~}}

\newcommand{\powheg}    {{\sc{powheg}~}}

\usepackage{array, graphicx, subfigure}
\usepackage{graphics}
\begin{document}
\title{Extracting Muon Momentum Scale Corrections }
%\title{A Simple Method to Determine Muon Momentum Scale and Misalignment Corrections }
\subtitle{for Hadron Collider Experiments}
%\author{A. Bodek and U. Sarica}
%\institute{Department of Physics and Astronomy, University of
%Rochester, Rochester, NY  14627-0171}

\author{A. Bodek\inst{1}, A.  van Dyne\inst{1}, J. Y.  Han\inst{1}, W. Sakumoto\inst{1} and A. Strelnikov\inst{1}  }
\institute{Department of Physics and Astronomy, University of
Rochester, Rochester, NY  14627-0171,  USA}

\date{Received: date / Revised version: date  Sept. 25, 2012}
% $\nu_\mu, \nub_\mu$
\abstract{
We present a simple method for the extraction of corrections for bias in the measurement
of  the momentum of muons in hadron collider experiments.  Such bias can originate from 
a variety of  sources such as  detector misalignment, software reconstruction bias, and uncertainties
in the magnetic field.
The two step method uses the mean 
$\langle 1/p^\mu_T \rangle$
for muons from $Z\to \mu\mu$ decays to determine the momentum scale corrections in bins of charge, $\eta$ and $\phi$. In the second step, the corrections are
tuned by using the  average invariant mass  $\langle M^Z_{\mu\mu}\rangle$
of $Z\to \mu\mu$ events
 in the same bins of charge $\eta$ and $\phi$.
 The forward-backward asymmetry of $Z/\gamma^{*} \to \mu\mu$ pairs as a function of $\mu^+\mu^-$ mass, and the $\phi$ distribution of $Z$ bosons in the Collins-Soper frame are used to ascertain that  the corrections remove the bias
 in the momentum measurements for positive versus negatively charged muons. By taking the sum and difference of the momentum scale corrections for positive and negative muons, we isolate additive
 corrections  to $1/p^\mu_T$ that  may originate from  misalignments and multiplicative corrections that
 may originate from mis-modeling of the magnetic field $(\int \vec{B} \cdot d\vec{L})$.  This method has recently been used in the CDF experiment at Fermilab and in the CMS experiment at the Large Hadron Collider at CERN.
\PACS 
  {  
      {20.26}{Experimental methods and instrumentation for elementary-particle and nuclear physics}  
      % \and
      %         {13.15.+g}{	Neutrino interactions} 
                                    } % end of PACS codes
} %end of abstract
\maketitle
%
% Section 1

\section{Introduction}
\label{intro}

In general, the reconstruction of the momentum of muons in hadron collider experiments
(e.g.. CDF, D0, ATLAS, CMS) is biased.  Bias originates from detector misalignments,
the reconstruction software, and uncertainties in the magnetic field    %(BdL).
$(\int \vec{B} \cdot d\vec{L})$.
%Even for Monte Carlo (MC) generated events, which in principle have no detector
%misalignment, it has been found that the reconstruction software may be biased.
Monte Carlo (MC) generated events start with no biases, but inaccurate inputs for
the detector aligmment, magnetic field, and running conditions
%during detector simulation
can induce biases during the reconstruction of the events.
The bias in  the
%reconstruction of the 
reconstructed momentum of muons depends on the charge of the muon, and on the
$\eta$ and $\phi$ coordinates\cite{cms_coord} of the muon track. The bias in the
reconstruction of the muon momentum in the data
and in the   %MC generated 
simulated events is not necessarily the same. Therefore, comparison of data and
reconstructed MC events require the removal of the bias from both data and reconstructed
MC samples.

 Precision measurements such as the  charge asymmetry in the production
 of $W$~bosons,  the measurement of the forward-backward
 asymmetry ($A_{fb}$) of  $Z/\gamma^{*} \to \mu\mu$  events as a function of 
 the$\mu^+\mu^-$ mass, measurements of
  angular distributions, and searches for new high mass states decaying to $\mu^+\mu^-$ pairs 
  are very sensitive to bias in the measurement of muon momenta. 
   The  reconstruction bias also worsens the detector resolution since it  depends on
  the charge of the muon and the $\eta$ and $\phi$ coordinates of the muon track.
 % and the mass of the Higgs boson mass in the {$H \rightarrow$ 4-lepton} channel 
%are compromised 
%by detector misalignments which worsen the  resolution and introduce bias in the measurement of muon momenta. 
In this paper, we present a simple data-driven method for the extraction of misalignment
and muon scale corrections  from  the $Z/\gamma^{*} \to \mu\mu$ event samples. 
The paper is organized as follows.  In section \ref{method} we present a general overall view of the method.  We then follow with additional details and application of the method to real collider data.

\subsection{The Method}
\label{method}

Since the $Z$ mass is well known, $Z/\gamma^{*} \to \mu\mu$ events have been
previously used to check on the momentum scale of reconstructed muons.
The difficulty is that the $\mu^+$ and $\mu^-$ are correlated, and the mass of
the final state depends on the momentum of the two muons.  Therefore, we devise
a two step process as described below.
 
In the first step, we obtain initial corrections in bins of charge,  $\eta$ and $\phi$.
These corrections are uncorrelated, remove all the bias, and yield the correct average
mass of the $Z$ boson for the sample.  In the second step we fine-tune the corrections
using the mass of the $Z$ boson for each bin of charge (Q),   $\eta$ and $\phi$. 

\subsubsection{Monte Carlo sample for a perfectly aligned detector}
\label{perfect}
%Since the reconstruction of the muon momenta are biased in both  data and MC, 
We begin by constructing a MC sample of  $Z/\gamma^{*} \to \mu\mu$ events for a perfectly
aligned and unbiased detector as follows.  
We start with a      %MC
simulated sample of $Z/\gamma^{*} \to \mu\mu$ events with identical selection cuts as
the data.  We know that the reconstruction of the momentum for these MC events may be biased.
Therefore, instead of using the reconstructed MC information, we use the  generated momentum
and smear it with a functional form that represents  the experimental
resolution as function of $\eta$. This process yields a sample of $Z/\gamma^{*} \to \mu\mu$
events for a perfectly aligned detector. 

For the Tevatron, we use \pythia \cite{pythia} for the generated sample and weight the transverse momentum and rapidity distributions \cite{cms_coord} by correction factors to bring the distributions into agreement with published\cite{dsdy} CDF data. 

For the LHC, we use \powheg \cite{powheg} for the generated sample  and weight the
transverse momentum and rapidity
distributions of the $Z$ bosons by a correction factor to bring them into agreement
with published\cite{dsdpt} CMS data.

 \subsubsection{Summary of the first step}
In the first step, we obtain initial momentum scale corrections in  bins
of charge $(Q)$, $\eta$ and $\phi$ by requiring that 
the average of $1/p^\mu_{T}$ ($\langle 1/p^\mu_{T}\rangle$) of selected muons from $Z$ decays
for data and reconstructed MC to be the same as that for the perfectly aligned sample.   
This yields a  lookup table of momentum scale corrections for 
$\mu^+$ and $\mu^-$ events (separately)  for both
data and reconstructed MC. 
Since we use a lookup table,  we are not constrained by a particular
 functional form for the parametrization of the  corrections. 
 These corrections remove all bias in the reconstructed momenta. 
 
We refer to this step as the $\langle 1/p^\mu_{T}\rangle$ based corrections.
Since we only use the $\langle 1/p^\mu_{T}\rangle$ for
individual muons, the correction for each bin  in $\eta / \phi$ 
is  uncorrelated with the correction for any of the other  $\eta / \phi$ bins.
Since the corrected  $\langle 1/p^\mu_{T}\rangle$ for each bin is now the same as
the $\langle 1/p^\mu_{T}\rangle$ for the perfectly aligned and unbiased MC sample,
the average mass of the $Z/\gamma^{*} \to \mu\mu$ 
  is also correct.  When this procedure is applied to the sample of  reconstructed MC events,
 we find that  these first step corrections remove all the biases in the  reconstructed momentum for all bins
 in $Q$,  $\eta$ and $\phi$.

 \subsubsection{Summary of the second step}
 
 When we extract these  first step corrections for the data, we  assume that the perfectly aligned MC sample
correctly models the rapidity and transverse momentum  distributions for  the production and decay of $Z/\gamma^{*} \to \mu\mu$ events,  including final state radiation of
photons.   In addition, we assume that the MC correctly models  the detector acceptance and efficiencies.
These assumptions are correct on average because the rapidity and transverse momentum 
distributions for the MC is usually tuned to describe the data.
Similarly, the efficiency for the reconstruction of muons as a function of
%transverse momentum 
the muon $\eta$
is also extracted from the data.
Nonetheless, when we apply this procedure to the data, we find that although the average mass of $Z\to \mu^+\mu^-$ events
is correct, we see  random scatter in the average $Z$ mass for different  $\eta / \phi$  bins in data which is not seen in  the corrected reconstructed MC sample. 
This random scatter is due to the fact that there are variations in the muon detection
efficiency
%(which is a function of $p^\mu_{T}$)
for the different $\eta / \phi$ bins, which is not perfectly modeled in the MC.    

 In order to be independent of modeling of detector efficiencies, and also be independent of  the modeling
 assumptions for the   production of $Z/\gamma^{*} \to \mu\mu$ events as a function of rapidity and transverse momentum,  we fine-tune the corrections by requiring that the reconstructed $Z$ mass is the same as for the perfectly aligned detector for  $\mu^+$ and $\mu^-$ events  in each bin in $\eta$ and $\phi$.  This removes the scatter in the average $Z$ mass for different  $\eta / \phi$  bins. We refer to this step as the $\Delta M/M$ tuning.   
 
Next, by taking the sum and difference of the  $\mu^+$ and $\mu^-$ momentum scale corrections,
we extract {\em additive} (in $1/p^\mu_{T}$)  corrections  that are caused by  misalignments,
and {\em multiplicative} (in $1/p^\mu_{T}$) corrections that are caused by mis-modeling of the
magnetic field (or from  mis-modeling of the integral of       %BdL
$\vec{B} \cdot d\vec{L}$) as a function of $\eta$ and $\phi$. 
 
 A detailed description of the method is given below.

\section{Data Set and Event Selection}

For CMS,
we use $Z \to \mu\mu$ events generated by the POWHEG Monte Carlo\cite{powheg}
%of $Z \to \mu\mu$ interfaced to PHYTIA
followed by PYTHIA which models
parton showering and final state radiation.  We apply event weighting corrections
which are a function of $Z$ transverse momentum ($P^Z_T$) and rapidity ($y$)
%the rapidity and transverse momentum distributions of the MC
%such that it describes the data.
to ensure that the transverse momentum
and rapidity distributions in  MC match  the data.

 For both data and reconstructed MC events
 we require  both muons to be isolated.  In the definition of isolation for a muon 
we use information only from the track and hadron calorimeter. 
   If the electromagnetic (EM) calorimeter energy is not included in the isolation
requirement, the  momentum  dependence of the efficiency is expected to be constant. 
If the  EM energy is included in the isolation requirement, 
 then photons from final state radiation result in a momentum dependence of the efficiency, and also in 
 a complicated correlation between  the efficiency of the two muons. 
 
 For example, for the CMS detector we use the following selection requirements: 
 
\begin{itemize}
\item  $p^{\mu}_T>25$  GeV/c
     %(to ensure high muon trigger efficiency)
       on the muon with the largest $p_T$ to ensure high muon trigger efficiency,
       and $p^{\mu}_T>20$  GeV/c for
       the second muon.
\item Detector $|\eta|<2.4$       %(to fall with the acceptance of the tracker)
                            (tracker acceptance)
\item Mass selection: $60 < M_{\mu\mu}< 120$~GeV/$c^2$ for the $1/p_T$ based correction
\item $86.5 < M_{\mu\mu}< 96.5$~GeV/c$^2$ for cross checks on the $\Delta M/M$ tuning. 
\end{itemize}

%The $<1/p^\mu_{T}>$ based correction requires 
%that the Monte Carlo that is used is tuned to match the data as closely as possible.
%For example, the  muon reconstruction efficiency  as a function of $\eta$ is extracted from the  data,
%and the efficiency scale factor obtained from the data is  applied to the MC to correct for the difference
 % of the efficiency between  data and MC.
  % However, although the efficiency correction
 % is correct on average, it may not account for
 % a possible $\phi$ dependence, or a possible $p_T$ dependent wiggles
%  of the efficiencies for  muon $p_T>$ 25 GeV (which may be different for different $\eta$ and $\phi$ bins). 
 % This is the reason that we use $\Delta M/M$ to fine tune  the corrections.
 % With the  $\Delta M/M$ tuning, the corrections become independent of
%  the  modeling of detector efficiencies. 

\section{Reference Plots Used in the  Muon Momentum Study} \label{sec:reference_plot}

A misalignment of the tracker  generates distortions in several  kinematic distributions of Drell-Yan ($Z/\gamma^{*} \to \mu\mu$) events in the $Z$ boson mass region. 

In our analysis we use several kinematic distributions, including the invariant mass of the
dimuon pair ($M_{\mu\mu}$),  the angles $\theta_{CS}$ and $\phi_{CS}$ of the negatively charged
muon in the
Collins-Soper frame ~\cite{collins}, and the forward-backward asymmetry of the negatively
charged muon in the Collins-Soper frame. 

The Collins-Soper \ frame is the rest frame of the dilepton pair.   In this frame, we define the  momentum vector of the beam particle as $\vec {P_A}$ and the momentum vector of the target particle as $\vec{P_B}$.  For proton-antiproton collisions (e.g. Tevatron)  the beam particle is defined as the proton and the target particle is defined as the antiproton.
The z-axis bisects the beam particle direction
and the opposite of the
target particle direction in the dilepton rest frame. 
The positive $z$ axis is along the beam particle direction. 
For proton-proton collisions (e.g. LHC),  the beam particle (i.e. positive $z$ axis) is defined  as the proton beam that points in the direction of the rapidity of the dilepton pair. 

The  angles $\theta_{CS}$ and $\phi_{CS}$ are defined \cite{collins}  by 
\begin{eqnarray}
  \small
  \cos \theta_{CS}& =& \frac{2}{M_{\mu\mu}\sqrt{M_{\mu\mu}^{2} + P_{\rm T}^{2}}} (p_{1}^{+}p_{2}^{-} - p_{1}^{-}p_{2}^{+})\nonumber \\
    \tan \phi_{CS} &=& \frac{\sqrt{M^{2}_{\mu\mu}+P_{\rm T}^{2}}}
   {M_{\mu\mu}} \cdot \frac{\vec{\Delta}_{r}\cdot \hat{R}_{T}}{\vec{\Delta}_{r}\cdot \hat{P}_{\rm T}}
    % \frac {M\sqrt{M_{\mu\mu}^{2} + P_{\rm T}^{2}}}{M_{\mu\mu} 
 %(p_{1}^{+}p_{2}^{-} - p_{1}^{-}p_{2}^{+})
  \label{CScost}
\end{eqnarray}
Here,  $p_{1}$ and $p_{2}$ are the 
four-momentum of negatively and positively charged muons, respecctively,  
$P_{\rm T}$ is the transverse momentum of the dimuon pair in the laboratory system,  
 $p^{\pm}$ corresponds to $\frac{1}{\sqrt{2}}(p^{0}\pm p^{3})$, 
$\Delta^{j}=p_1^{j}-p_2^{j}$, $\hat{P_{\rm T}}$ is a transverse unit vector in the direction of $\vec{P_{\rm T}}$,
  and $\hat{R_{T}}$ is a transverse unit vector in the direction of $\vec{P_{A}}\times \vec{P_{\rm T}}$.

We  define $\phi_{CS}$ to be the angle between the direction of the
 $Z/\gamma^{*}$ boson $p_T$  and the direction of the negatively charged lepton,

When integrated over all  $\phi_{cs}$ the differential cross section can be written as :
\begin{eqnarray}
\frac{d{\sigma}}{d\cos{\theta}} \propto (1+\cos^2{\theta})
+ \frac{1}{2}A_0(1-3\cos^2{\theta}) + A_4\cos{\theta}
\label{AngleFunc3}
\end{eqnarray}
where  $A_0 (M_{\mu\mu}, y, p_T)$ originates from QCD gluon radiation and  $A_4(M_{\mu\mu}, y, p_T)$
originates from electroweak interference. 
The forward  (f) backward (b)  asymmetry in the Collins-Soper frame is defined as.
\begin{eqnarray}
A_{fb} & =&  \frac{N_{f}-N_{b}}   {N_{f}+N_{b}} 
\end{eqnarray} 
where $N_{f}$ and $N_{b}$ are the number of events for positive and negative $\cos\theta_{CS}$,
respectively.  For a detector with 100\% acceptance over all  $\cos\theta_{CS}$ the
forward-backward asymmetry  is given by $A_{fb}= \frac{3}{8}A_4$.

Since the misalignments in data and MC are different, the distributions are distorted in different ways for data and MC. 
Detector misalignments may be responsible for the following:

\begin{itemize}

 \item Cause a
 charge, $\eta$, and $\phi$ dependent bias in the measurement of the muon momentum which also
 worsens the resolution.
 
\item   Cause a  charge, $\eta$, and $\phi$ dependence of  the average reconstructed $Z$ boson mass. 
% Note that the  expected
%$Z$ boson mass is known from the generated (post FSR with EW interference and smeared
%by the experimental resolution ) spectrum in MC for accepted events. 

\item Distort and widen the  overall shape of the  $Z/\gamma^{*} \to \mu\mu$ mass distributions.
%in the $Z$ mass region.  
%between the data and MC 
 %(if data and MC have different misalignments or different resolutions).

\item A charge dependence in the reconstructed muon momentum creates unphysical
      wiggles in the forward and backward lepton angle asymmetry ($A_{fb}$)
      in the Collins-Soper~\cite{collins} (CS) dilepton rest frame for Drell-Yan events
      as a function of dilepton mass (in the region of the  $Z$  peak). This yields
      one of two powerful checks on a difference in the momentum scale between
      positive and negative muons.
\item For low $p_{T}$ $Z/\gamma^{*}$ bosons ($p^Z_{T}<10$~GeV/c),  the $\phi$
      distribution in the CS frame ($\phi_{CS}$) is expected to be flat.
  %Here   $\phi_{CS}$  is defined as  the angle between the
  %direction of the  $Z\gamma^{*}$ boson $p_T$  and the direction of the positive lepton.
      However, since we  define $\phi_{CS}$ to be the angle between the direction of the
      $Z/\gamma^{*}$ boson $p_T$  and the direction of the negatively charged lepton, 
      resolution smearing in the measurement of the  muon momentum results in an
      excess of events near $\phi_{CS} = 0$ and $\pm \pi$ in the reconstructed $\phi_{CS}$
      distribution.
      The level of the excess at $\phi_{CS} = 0$ and $\pm \pi$ is expected to be the same
      if the muon momentum scales and resolutions are the same between $\mu^{+}$ and $\mu^{-}$. 
      For $Z/\gamma^{*}$ events with $p^Z_T=0$  there is no preferred $x$ axis. However, if
      there is a difference in the reconstruction bias for positive and negative muons,
      events which are produced with $p^Z_T=0$ are reconstructed with  $p^Z_T$  along either
      the positive  ($\phi_{CS} = 0$)  or the negative muon  ($\phi_{CS}=\pm \pi$)  direction.
      Therefore, the $\phi_{CS}$ distribution in the low $p^Z_{T}$ region provides the second
      powerful check on a difference in the momentum scale between positive and negative muons.
  \end{itemize}
  
In our study we use the following two kinematic distributions as reference plots to test the validity of the momentum
corrections. These reference plots are not used in the extraction of the momentum corrections. They are only
used to ascertain that the correction factors actually work.
\begin{itemize}
\item  $A_{fb}$  for $Z/\gamma^{*} \to \mu\mu$ events as a function of mass.
\item  $\phi_{CS}$  in two  $Z/\gamma^{*}$ $p_T$ bins:  $0<p^Z_{T}<5~$GeV/c,
       and  $5<p^Z_{T}<10$~GeV/c.
\end{itemize}

The following distributions are used to determine the momentum correction factors
and also determine the $\eta/\phi$ dependence of the momentum resolution:
\begin{itemize}
\item  The $1/p^\mu_{T}$ distributions for positive and negative muons in bins of $\eta$ and $\phi$.

\item The overall  di$\-$muon invariant mass spectrum ($M_{\mu^{+}\mu^{-}}$).
\item The average $Z/\gamma^{*}$ mass
    %  (in the Z peak region,
    %  $86.5 < M_{\mu\mu} < 96.5$ GeV/$c^{2}$)  as a function of  $\phi$ of the
    %  $\mu^+$, or the   $\mu^-$.
       in the Z peak region as a function of $\eta$ and $\phi$ of the $\mu^+$ or the $\mu^-$.
       If the $\mu^+$ of the pair is binned in $\eta$ and $\phi$, its partner is allowed to
       be in any bin, and vice versa.
       A broad window of 60--120 GeV/$c^{2}$ is used for initial tuning, and a tighter
       window of 86.5--96.5 GeV/$c^{2}$ is used for the final tuning. 
%(and also the mass with no sign requirement).
%
%\item The average $Z$ mass (in the Z peak region, $86.5 < M_{\mu\mu} < 96.5$ GeV/$c^{2}$)  as a function of  $\eta$ of  the  $\mu^+$, or the  $\mu^-$  (and also the mass with no sign requirement.)
%
\end{itemize}
We use the same procedure to extract the corrections for data and reconstructed MC.  Since for the MC we know the generated muon momentum, we can use the generated information in the MC sample as an additional check on the procedure.
%Figure 1
\begin{figure}
%\begin{center}
\includegraphics[width=3.4in,height=3.5in]{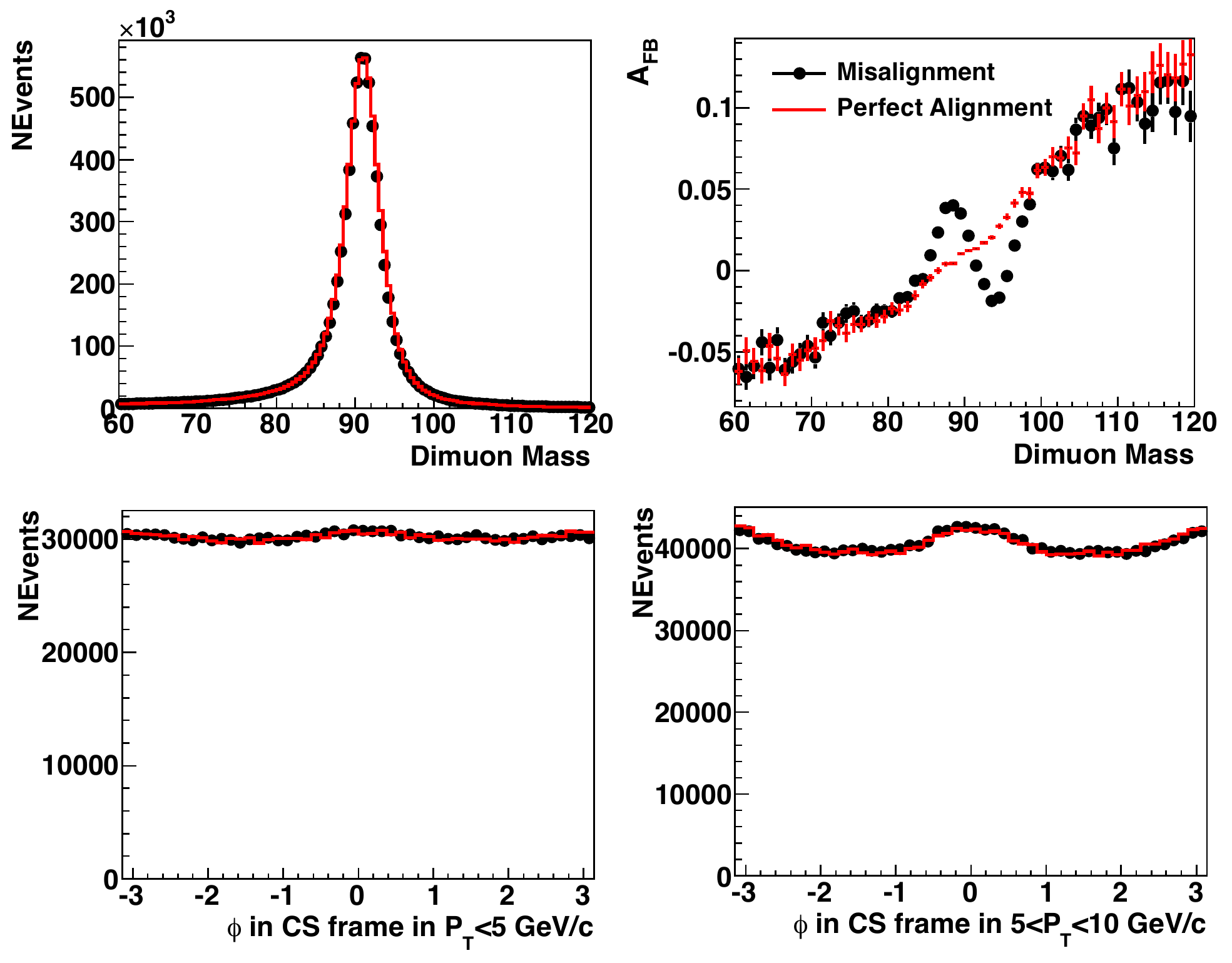}
\caption{  
 An example of the first set of  reference plots ($M_{\mu^{+}\mu^{-}}$, $A_{fb}$, and $\phi_{CS}$)
 for a CMS-like detector at the LHC for 7 TeV in the center of mass. The red histograms are the distributions for a perfectly aligned detector and the black points are for one example of a misaligned detector. 
  The kinematic selection cuts are:  muon $p^\mu_{T}>20$ GeV/c and $|\eta|<2.4$ for both muons and $60<M_{\mu\mu}<120$~GeV/c$^{2}$.  
Top Plots:  The $\mu^+\mu^-$ invariant mass distribution (left)  and $A_{fb}$ (right). 
 Bottom plots:  The $\phi$ distribution in the Collins-Soper frame in boson $p^Z_{T}<5~GeV/c$ (left) and $\phi$ in the Collins-Soper frame in boson $5<p^Z_{T}<10~GeV/c$ (right)  distributions. (Color online).} 
 \label{fig:ref1_gen_etasm}
%\end{center}
\end{figure}
% FIG  2
\begin{figure}
%\begin{center}
  \includegraphics[width=3.4in,height=3.5 in]{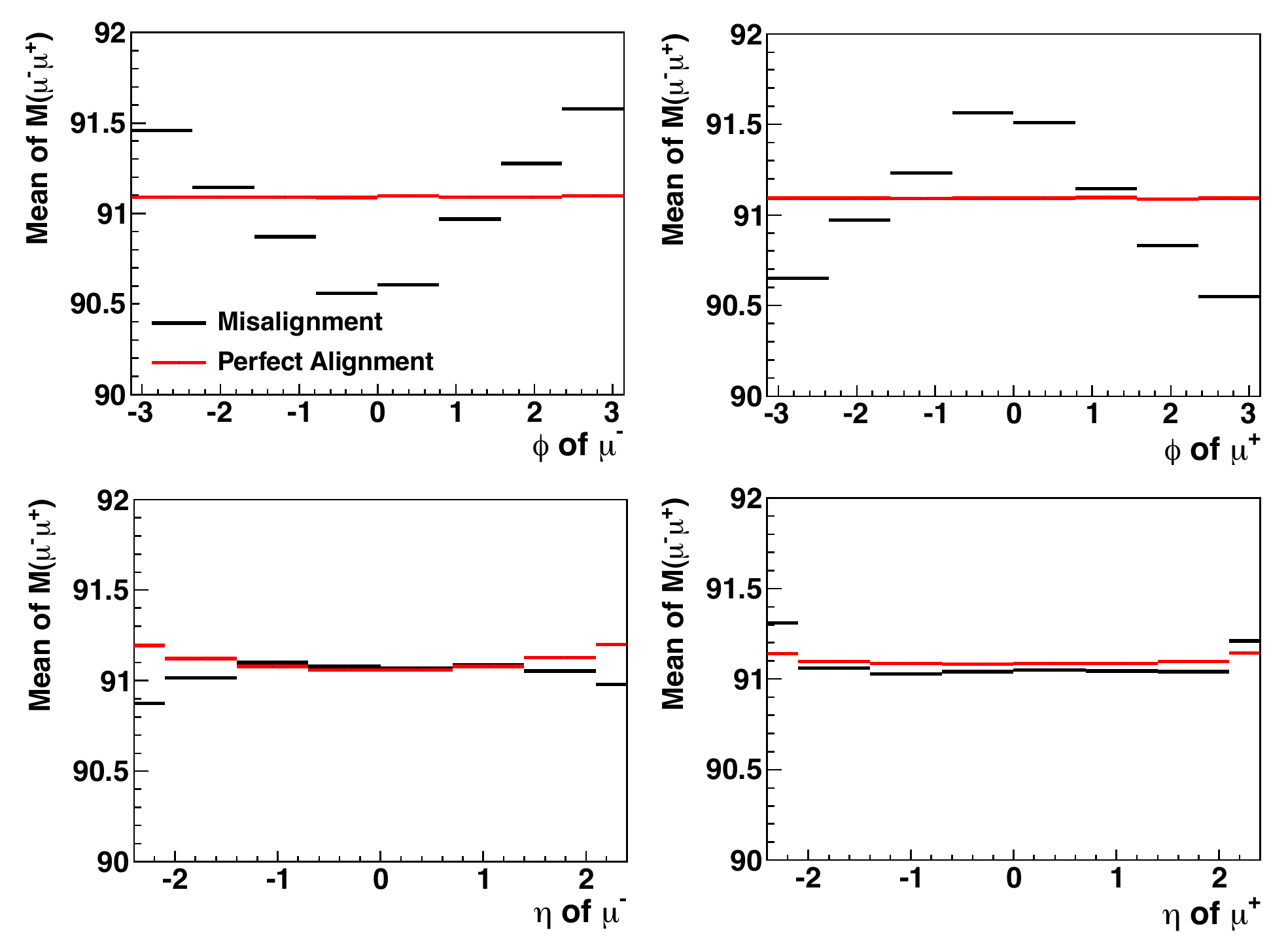}
\caption{An example of the second set of  reference plots  for a CMS-like detector at the LHC for 7 TeV in the center of mass.  The red histograms are the distributions for a perfectly aligned detector and the black points are for one example of a misaligned detector. 
 Shown are the  reference plots for average $Z$ mass  ($86.5<M_{\mu\mu}<96.5~GeV/c^{2}$) as a function of 
 $\phi$ (top) or $\eta$ (bottom) of the $\mu^+$ and $\mu^-$.
  The kinematic selection cuts are:   muon $p_{T}>20$ GeV/c and $|\eta|<2.4$ for both muons. 
}
 \label{fig:ref2_gen_etasm}
%\end{center}
\end{figure}
  Fig.  \ref{fig:ref1_gen_etasm} and  \ref{fig:ref2_gen_etasm} show examples of reference plots  for a perfectly aligned MC for a  CMS-like detector  for proton-proton collisions at 7 TeV in the center of mass (red histograms). Also shown are the same reference plots for one example of a misaligned detector (black points).
  Only generated information was used to produce these sample reference plots. For purpose of illustration we have assumed  100\% efficiency and a CMS-like momentum resolution.   
 The following kinematic selection cuts were applied:  $60<M_{\mu\mu}<120$~GeV/c$^{2}$,  muon $p^{\mu}_{T}>20$ GeV/c and  $|\eta|<2.4$ for both muons.  In an actual application, the reference plots  should also include the effect of  detector efficiency and  geometrical cuts which are specific to the experiment. 
 
  The reference plots  which are shown in Fig.  \ref{fig:ref1_gen_etasm} and  \ref{fig:ref2_gen_etasm} are 
 the $M_{\mu\mu}$ distribution,   $A_{\rm fb}$ versus $M_{\mu\mu}$, 
the distributions in  $\phi_{cs}$ for $P_T^Z<5$ GeV/c, and for  $5< P_T^Z<10$ GeV/c, and
the average   $Z$ mass ($86.5<M_{\mu\mu}<96.5$~GeV/c$^{2}$) versus muon $\eta$ and $\phi$ for positive and negative  muons.  (Note that the  the very  small QCD, EW (diboson), $\tau^+ \tau^-$ and top-antitop background is also included in the distribtributions).

 %which (as mentioned previously) are obtained  from the generated spectrum  (with $p^Z_T$ tuning)  
   %including the  the effects of FSR, resolution smearing, acceptance, efficiencies, small
  % backgrounds,   and  detector $\eta$ and $p^\mu_T$   requirements, 
 
 The  reference  plots   for the  reconstructed data (and reconstructed MC)  should be in agreement with  these 
 perfect alignment reference plots after all the momentum scale corrections
 are applied.

\section{Muon Momentum Correction (step 1):  $\langle 1/p^\mu_T\rangle$ based corrections}

\begin{figure}
%\begin{center}
\includegraphics[width=3.2in,height=2.5in]{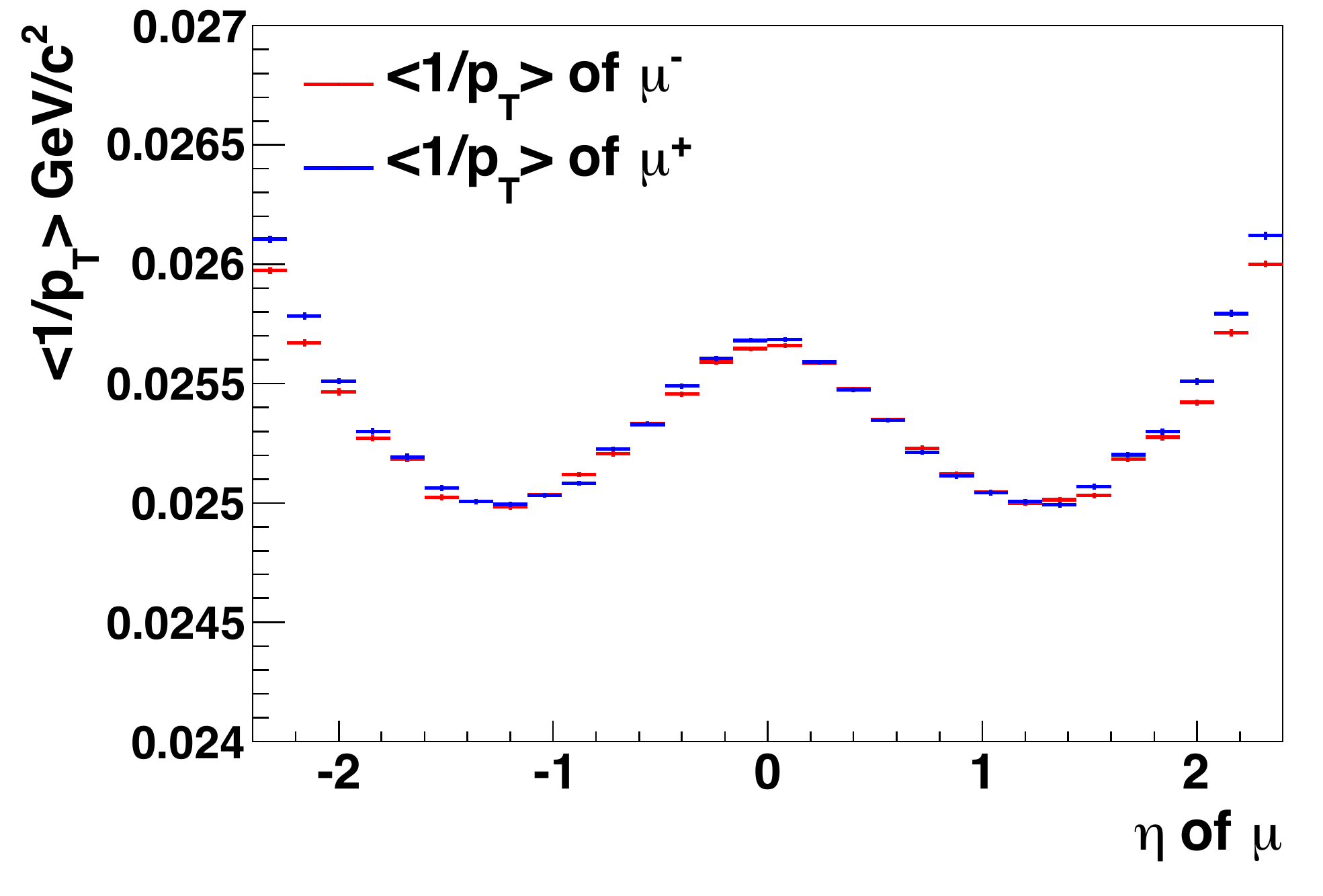}
\caption{  The mean  $\langle 1/p^\mu_{T}\rangle$ for a perfectly aligned detector
for proton-proton collisions at 7 TeV in the center of mass.    Note that because of the  $p^\mu_{T} > $ 25 GeV/c requirement,  the mean  $\langle 1/p^\mu_{T}\rangle$  is a function of  $\eta$.
  At large $\eta$, because of  electroweak interference,
the mean  $\langle 1/p^\mu_{T}\rangle$  for positive muons (shown in blue) and negative muons
(shown in red) is different.
} 
 \label{fig:ref3_gen_etasm}
%\end{center}
\end{figure}

The correction factor  $C^{Data/MC} (Q, \eta, \phi)$, is defined as the  difference in the mean
$\langle 1/p^\mu_{T}\rangle$ between the mean  $\langle 1/p^\mu_{T}\rangle$
for an ideal  perfectly aligned MC  and
reconstructed data (or reconstructed MC).
Since the $\langle 1/p^\mu_{T}\rangle$ based muon momentum correction factors are obtained in the
range of $p^\mu_{T} > 25$ GeV/c, the corrections are iterated to account for the fixed
$p^\mu_{T} >25$ GeV/c until the mean $\langle 1/p^\mu_{T}\rangle$ of muons
in the corrected data (or corrected reconstructed MC) agree with the mean $\langle 1/p^\mu_{T}\rangle$
of the perfectly aligned MC.  Note that because of the  $p^\mu_{T} > 25$ GeV/c requirement,  the mean  $\langle 1/p^\mu_{T}\rangle$  is a function of  $\eta$, as shown in Fig. \ref{fig:ref3_gen_etasm}.  
At large $\eta$, because of  electroweak interference,
the mean  $\langle 1/p^\mu_{T}\rangle$ is  different for positive muons which are shown in blue and and negative muons which are shown in red.
%
%
 %Figure 3

 %Since the $Z$ mass is known, and the $p^Z_T$ spectrum
%in MC has  been tuned to agree with the data, this procedure in principle should yield an absolute calibration of %the momentum scale for each $\eta$, and $\phi$ bin.

In general, an overall momentum scale (e.g. error in the $B$ field) should be the same for
positive and negative muons. A misalignment would results in a difference in the mean
$\langle 1/p^\mu_{T}\rangle$ 
between positive and negative muon. A portion of the muon momentum correction 
that corrects for a misalignment is additive in 1/$p^\mu_{T}$.
A portion of the muon momentum correction that corrects an inaccurate        %BdL
$\int \vec{B} \cdot d\vec{L}$
is multiplicative in 1/$p^\mu_{T}$, and is the same for positive and negative muons.  

Therefore, the correction factors $C^{Data/MC} (Q, \eta, \phi)$ for positive and negative
muons are then regrouped to form two different corrections,
\begin{itemize}
 \item A muon momentum scale multiplicative correction ($D_{m}$) that could originate
 from an incorrect  integral of      %B*dL.
 $\vec{B} \cdot d\vec{L}$.
 \item An additive correction for the bias ($D_{a}$) that could originates from misalignment.
\end {itemize}

% We define the multiplicative correction as:
%  $$D_{m} = (C^{Data/MC} (+, \eta, \phi) + C^{Data/MC} (-, \eta, \phi))/2.0$$
%   and the additive correction as: 
% $$D_{a}=(C^{Data/MC} (+, \eta, \phi) - C^{Data/MC} (-, \eta, \phi))/2.0$$

 In addition, we define an overall scale correction $G$ which is determined by the known $Z$ mass
 peak position. After the momentum scale corrections, we expect to obtain $G=1.0$.  
In the equations below we refer to
the perfectly aligned resolution smeared MC  as $MC(gen)$, and $MC(rec)$ denotes the  MC at  the reconstructed (misaligned)  level.

\begin{eqnarray*}
 \lefteqn{C^{Data/MC} (Q, \eta, \phi)  =   } \nonumber \\
 &  \langle 1/p_{T}^{MC(gen)}      (Q, \eta, \phi)\rangle  - 
    \langle 1/p_{T}^{Data/MC(rec)} (Q, \eta, \phi)\rangle \\
   \nonumber \\
  & D_{m} (\eta, \phi) = (C^{Data/MC} (+, \eta, \phi) + C^{Data/MC} (-, \eta, \phi))/2 \nonumber\\
  & D_{a} (\eta, \phi) = (C^{Data/MC} (+, \eta, \phi) - C^{Data/MC} (-, \eta, \phi))/2 \nonumber\\
 % &  \frac{1}{p_{T,\eta,\phi:corrected}^{\pm}} = 
 %    \frac{1}{p^{\pm}_{T}}  \times ( 1 + D_{m}(\eta, \phi)/\langle 1/p^{\pm}_{T}\rangle )
 %    \pm D_{a}(\eta, \phi) \nonumber \\
       & \frac{1}{p_{T,\eta,\phi:corrected}^{\pm}} = 
     \frac{1}{p^{\pm}_{T}}  \times M (\eta, \phi)   \pm A (\eta, \phi) \nonumber \\
&  M (\eta, \phi) = 
1 +\frac{ 2D_{m}(\eta, \phi)}{\langle 1/p^{+}_{T}\rangle + \langle 1/p^{-}_{T}\rangle  } \nonumber\\
&  A (\eta, \phi)   = D_{a}(\eta, \phi) -\frac{ D_{m}(\eta, \phi)(\langle 1/p^{+}_{T}\rangle - \langle 1/p^{-}_{T}\rangle) }{\langle 1/p^{+}_{T}\rangle + \langle 1/p^{-}_{T}\rangle  }  \nonumber \\
&   {p_{T,scale+\eta,\phi:corrected}^{\pm}}=G \times  {p_{T,\eta,\phi:corrected}^{\pm}} \nonumber \\
\end{eqnarray*}
Here, $C^{Data/MC}$ is the muon momentum correction factor for the data or reconstructed  MC
in  bins of $Q$, $\eta$, and $\phi$  of the muon (e.g. $8\times 8$ matrix in $\eta$ and $\phi$
for each muon polarity). 
This $\langle 1/p^\mu_{T}\rangle$ correction corrects for  the charge, $\eta$, and $\phi$
dependence
of the mis-reconstructed momentum, as well as an overall scale to yield the correct $Z$ mass.  

After the application of the multiplicative and additive corrections,
the $Z$ peak position at the reconstructed level in data and MC is tuned
with a multiplicative corrections $G^{data}$ and  $G^{MC}$
  (which are expected to be close to 1.0)   to
  agree with that of the perfectly aligned MC. 
  We chose to define the peak position  by fitting the generated spectrum (post FSR)  in a narrow $Z$ mass region  (88 to 94 GeV) to a  Breit-Wigner function.
   
In addition, we use the parameters  $\Delta$, and SF, to make sure that the resolution in the
reconstructed  Monte Carlo matches the resolution in data. 
Here, $\Delta$ and SF, are estimated by comparing the overall  $M_{\mu^{+}\mu^{-}}$
mass distributions between data and MC (using a $\chi^{2}$ test).   These parameters,
which are only applied to MC events,
%are defined by the following equations:
are used to tune the width of the  MC $M_{\mu^{+}\mu^{-}}$ distribution to match the data. This is done via
additional $p_{T}$ smearing:   
 \vspace{-0.2in}  
\begin{center}
\begin{eqnarray} % requires amsmath; align* for no eq. number
 %  p_{i}^{corrected} = p_{i} + T \times ( p_{i}^{gen.} - p_{i} ) \\
 %  \nonumber \\
  \frac{1}{p_{T}^{additional-smearing}} = \frac{1}{p_{T}} + \Delta \times
              %  Random::Gaus(1, SF)\nonumber
              {\cal N}(1,SF), \nonumber
\end{eqnarray}
\end{center}
%
% where $p_{i}$ is the reconstructed muon momentum in MC ($i=x$, $y$, and $z$) and $p_{i}^{gen.}$ 
% is the generated muon momentum in MC.  
where ${\cal N}(\mu, \sigma)$ is a random normal distribution with a mean of $\mu$ and
an rms of $\sigma$.

\section{Muon Momentum Correction (step 2):  further tuning using $\Delta M^Z$ } \label{sec:final_it}

  The  $\langle 1/p^\mu_{T}\rangle$  based corrections fully correct for
  $all$ reconstruction bias
  in the reconstructed Monte Carlo.  The average $Z$ mass in bins of $Q$, $\eta$ and $\phi$ for
  the reconstructed MC after the corrections is the same as for the perfectly aligned MC.

    % These efficiency variations introduce  residual 
     %scatter (but no bias) in the average $Z$ mass in bins of Q, $\eta$ and $\phi$ in the
   %  data after the corrections.  Therefore, apply additional tuning to the scale corrections
   %  as described below.
    
    We form the distributions of   $\frac {\Delta M^Z}{M^Z}$
    where $\Delta M^Z$ = $M^Z$(measured)-$M^Z$(expected) for all $\mu^+$ 
    and $\mu^-$  in $\eta/ \phi$ bins.
           We find that  after the 
      $\langle 1/p^\mu_{T}\rangle$ based corrections (step 1) 
     are applied, the bias in the measurement
between positive and negative muons is removed from both the data and reonstructed MC.  
   However, we find that there is scatter in  $\langle M^Z_{\mu\mu}\rangle$ spectra in the data 
   that is larger than in the reconstructed MC.  This scatter originates from a small  $\eta$ and $\phi$
    dependence in the trigger and reconstruction efficiencies in data that are not simulated
    perfectly in the MC.  Mis-modeling of the  transverse momentum
    dependence of the  efficiency for different $\eta$ and $\phi$
   yields an incorrect value of  $\langle 1/p^\mu_{T}\rangle$
 
      We correct for the additional  scatter in the data  by using the deviation in the 
     average invariant mass   $\Delta M^Z$ of $ \mu\mu$ events in each
      of the  $\eta / \phi$ bins for $\mu^+$ 
    and  $\mu^-$  to fine tune the momentum correction.
    
       For each $\eta$/$\phi$ bin in the data,  the value of $\Delta M^Z$ can be different from zero
      if the momentum scale for one of the muons in that bin ($p_1$) is sightly off. 
       % (because the efficiency is not modeled perfectly). 
       The fluctuations in the momentum scale
       for the other muon leg ($p_2$), which can end up in  any place
       in the detector, averages to zero. This is because after the application
        of the $\langle1/p^\mu_{T}\rangle$ based corrections, all biases are removed and the average
        of the momentum scale corrections for a large number of $\eta$ and $\phi$ bins is zero.
        
         The relation between $\frac {\Delta M}{M}$ and $\frac {\Delta p_\mu}{p_\mu}$ can be extracted from
   the following expressions:
        
         \vspace{-0.0in}  
%\begin{center}
\begin{eqnarray} % requires amsmath; align* for no eq. number
   M^2_{Z-Data} (Q, \eta, \phi) &= & 2 p_1 p_2 ( 1-\cos\theta ) \nonumber \\
     2\Delta M \times  M_{Z-Data} (Q, \eta, \phi) &= & \Delta p_1 \times (2 p_2 (1-\cos\theta )) \nonumber \\  
          2\Delta M \times  M_{Z-Data} (Q, \eta, \phi) &= & \frac {\Delta p_1}{p_1} \times (M_{Z-Data}^2) \nonumber\\
           2 \frac {\Delta M}{M}(Q, \eta, \phi) &= & \frac {\Delta p_1}{p_1} \nonumber 
          \end{eqnarray}   
%\end{center}
%
        Therefore, we fine tune the mean $\langle 1/p^\mu_{T}\rangle$ based correction
        by an additional factor  of $1+2 \frac {\Delta M}{M}(Q, \eta, \phi)$.
        We do this iteratively until the distribution for $\frac {\Delta M^Z}{M^Z}$ has
        the smallest rms about zero. 
       We refer to this step as the combined
         $\langle 1/p^\mu_{T}\rangle$ and $\Delta M$ based  correction.   
         
         In addition to
         removing bias, we find that for a CMS like detector the momentum resolution
           for 1 TeV muons is improved from $\pm8\%$ before the application
           of  momentum scale/alignment corrections
           to $\pm4\%$ after corrections.  In the measurement of the mass of
           125 GeV Higgs boson in the four muon channel, the systematic error in the
              mass scale  is reduced from 0.4\%  before the application
           of  momentum scale/alignment corrections to less than 0.1\% after corrections.

\section{Systematic Errors}

Once we fine tune the corrections using the  $\frac {\Delta M}{M}$ distributions, we find
that most of the systematic errors are removed.  The two step procedure is insensitive to the modeling
of the  efficiencies,  backgrounds,  or modeling of the rapidity and $p_T$ spectrum for the production of $Z/\gamma^*$ bosons. 
 As a test for the LHC samples, we removed the $p_T$ tuning from the generated MC sample 
and  did not subtract any of the backgrounds from data samples.
We repeated the entire process, and the resulting coefficients extracted from  the  combined 
$\langle 1/p^\mu_{T}\rangle$ and $\Delta M$ based  corrections
remained unchanged.

The errors in the momentum scale corrections originate primarily from the statistical
errors in the $\mu^+\mu^-$ sample.  The  samples are of $\approx$ 0.5 million  $\mu^+\mu^-$ events for
the Tevatron, and  a few million $\mu^+\mu^-$ events for the LHC,  respectively.

\section{Conclusion}

Precision measurements such as the  charge asymmetry in the production
 of $W$~bosons,  the measurement of the forward-backward
 asymmetry in $Z/\gamma^{*} \to \mu\mu$ events as a function of mass ($A_{fb}$), measurements of
  angular distributions, and searches for new high mass states decaying to $\mu^+\mu^-$ pairs 
  are very sensitive to bias in the measurement of muon momenta.

We presented a simple method for the extraction of corrections for bias in the measurement
of  the momentum of muons in hadron collider experiments.  Such a bias can originate from 
a variety of  sources such as  detector misalignment, software reconstruction bias, and uncertainties
in the magnetic field $(\int \vec{B} \cdot d\vec{L})$. The corrections are  obtained by using the average
$\langle 1/p^\mu_{T}\rangle$
of muons from $Z/\gamma^{*} \to \mu\mu$ events  in bins of  charge, $\eta$, and $\phi$ and
further tuned using the  $ \mu^+\mu^-$ invariant mass distributions.

% Corrections are extracted for both data
%and MC. 
%The muon momentum correction removes the bias from the charge (Q), $\eta$, and $\phi$ of the muon.

 The  $M_{\mu^{+}\mu^{-}}$, $A_{fb}$, and $\phi_{CS}$ distributions are used as reference plots to test
the procedure.
After the application of the combined $\langle 1/p^\mu_{T}\rangle$ and $\Delta M/M$
based  muon momentum correction, any reconstruction bias which may in general be 
a function of charge,  $\eta$, and $\phi$ is  completely removed.  All kinematic
distributions which are  used
as reference plots show good agreement with those expected for a perfectly aligned detector with
no reconstruction biases.

%=======================================================================================


\begin{thebibliography}{9}
\bibitem{cms_coord}
The coordinates in the laboratory
frame are ($\theta$,$\phi$,$z$), where $\theta$ is the polar angle relative to
the direction of one of the proton beams(the $+z$ axis), and $\phi$ the azimuth.
The pseudorapidity is $\eta=-ln~tan(\theta/2)$. For an $\mu^{+}+\mu^{-}$ pair
 $p_{T}=Psin\theta$,  $E_{T}=Esin\theta$, 
 the rapidity  $y=\frac{1}{2}ln\frac{E+p_{z}}{E-p_{z}}$, where
$P$ and $p_{z}$ are the magnitude and z component of the momentum,
and E is the energy of the $\mu^{+}+\mu^{-}$ pair.
\end{thebibliography}
 \end{document}